\documentclass[aps,prl,twocolumn,superscriptaddress,tight]{revtex4-2}

\usepackage{amssymb}
\usepackage{amsfonts}
\usepackage{color}
\usepackage{graphicx}
\usepackage{amsmath}
\usepackage{times}
\usepackage{bm}
\usepackage{float}
\usepackage{lipsum}
\usepackage{import}
\usepackage{enumerate}
\usepackage[colorlinks,linkcolor=blue,anchorcolor=blue,citecolor=blue,filecolor=blue,urlcolor=blue]{hyperref}

\hypersetup{colorlinks=true,linkcolor=blue,citecolor=blue,filecolor=blue,urlcolor=blue}

\newcommand{\SETI}{\ensuremath{\mathrm{SETI}}}
\newcommand{\SMsec}[1]{\href{run:supplement_shoucuo_0716.pdf}{\textcolor{blue}{SM Sec.~#1}}}

\begin{document}

\title{Beyond the Edge of Chaos: Stability--Expressivity Transfer in Reservoir Forecasting}

\author{Yao Du}
\affiliation{School of Physics and Information Technology, Shaanxi Normal University, Xi'an 710062, China}
\author{Xingang Wang}
\email{wangxg@snnu.edu.cn}
\affiliation{School of Physics and Information Technology, Shaanxi Normal University, Xi'an 710062, China}

\begin{abstract}
The edge-of-chaos heuristic has long served as a guiding principle for designing reservoir computers, yet its relevance to machine performance remains elusive. Here, taking the spectral radius of the reservoir network as the control parameter, we show that the radius yielding the best forecasting performance does not coincide with the Lyapunov edge of the isolated, teacher-forced, or closed-loop generative reservoir. By analyzing the collective dynamics of the teacher-forced reservoir, we find that the target dynamics are represented mainly by stable Lyapunov modes whose finite-time stability is strongly modulated by the input. This finding motivates a stability--expressivity transfer index, which balances the stability of these modes against their expressivity in representing the target. Across chaotic and quasiperiodic targets, and for both asymmetric and symmetric reservoirs, this index accurately identifies the optimal spectral radius for autonomous forecasting.
\end{abstract}

\maketitle

\emph{Background.---} Model-free prediction of chaos using reservoir computing (RC) has attracted considerable attention in nonlinear science~\cite{Maass2002,Jaeger2004,SJ2024,Lai2026}. A properly trained reservoir computer can not only accurately forecast the future evolution of an unknown system for several Lyapunov times, but also faithfully reproduce the long-term statistics of the target attractor~\cite{Pathak2017,Pathak2018,Lu2018,Fan2020,Gilpin2024,SisodiaJalan2026Crises}. This capability is remarkable because only a linear readout is trained, whereas the high-dimensional recurrent core, i.e., the reservoir network, remains fixed after construction. The performance of RC, however, is highly sensitive to the reservoir hyperparameters. Among them, the spectral radius is especially important, as it strongly affects the memory capacity, stability, echo-state property, and forecasting performance of the reservoir~\cite{Jiang2019,Carroll2020}. Although the optimal spectral radius can be found through direct parameter scans or well-developed optimization algorithms, such approaches require substantial computational resources and additional validation data, while providing limited dynamical insight. A question of interest in RC is therefore whether the optimal spectral radius can be determined from a dynamical principle rather than through exhaustive scans.

Inspired by findings that the computational capability of many complex systems is maximized near the boundary between ordered and chaotic dynamics, the edge-of-chaos heuristic has been adopted as a general dynamical principle for RC design and optimization~\cite{Kauffman1969,Langton1990,Sompolinsky1988,Bertschinger2004,Legenstein2007,Boedecker2012,Teuscher2022,Kobayashi2026}. The underlying intuition is a balance: excessive contraction suppresses the diversity of the response, whereas excessive instability destroys memory and reproducibility. The relevance of the edge-of-chaos heuristic to RC performance, however, remains unsettled. The best forecasting performance may persist over a finite interval of spectral-radius values rather than being concentrated at a single critical point~\cite{Jiang2019}; approaching the edge of stability does not generally optimize reservoir performance~\cite{Carroll2020}; and successful prediction can even be achieved with very low-connectivity reservoirs, including those with zero spectral radius and no recurrent links~\cite{Griffith2019}. In physical reservoirs of coupled phase oscillators, proper training can also induce target-dependent self-organization toward critical states~\cite{Wang2022Criticality}. These studies thus suggest that the forecasting optimum is not determined solely by a universal transition in the recurrent network. A more fundamental reason is that in RC, the reservoir does not operate in a single dynamical regime: the same recurrent core operates as an isolated autonomous network before training, as a teacher-forced system during learning, and as a closed-loop generative system during forecasting. These regimes generally have distinct Lyapunov spectra and, therefore, distinct stability boundaries. Consequently, the meaning of the edge of chaos is regime-dependent: the edge of the isolated reservoir characterizes the bare recurrent core, the edge of the teacher-forced reservoir characterizes input-driven stability, and the edge of the generative reservoir characterizes the trained closed-loop dynamics. None of these edges alone is guaranteed to select the spectral radius that yields the best autonomous forecasting. This raises the central question of the current study: if no single Lyapunov edge identifies the forecasting optimum, what dynamical criterion can identify it?

Here we address this question by resolving the finite-time Lyapunov spectrum of the teacher-forced reservoir. We first demonstrate, for chaotic and quasiperiodic targets and for both asymmetric and symmetric reservoirs, that the forecasting optimum is not predicted by the Lyapunov edge of any of the three operational regimes. We then show that target-induced responsiveness is concentrated in modes that remain stable on average while being strongly modulated in finite time. This observation leads to the introduction of the stability--expressivity transfer index, which identifies the forecasting-optimal spectral-radius regime from teacher-forced dynamics alone. The results reformulate reservoir optimization from searching for a single instability boundary to searching for a driven Lyapunov spectrum that supports expressive responses along stable directions.

\emph{Regime-dependent Lyapunov edges.---} We adopt the standard RC architecture for autonomous forecasting and consider three representative targets: the Lorenz oscillator~\cite{Lorenz1963}, the Chen oscillator~\cite{Chen1999}, and a quasiperiodic oscillator. During teacher-forced training, the reservoir is driven by the target signal and evolves as
\begin{equation}
\bm r_{n+1}=(1-\alpha)\bm r_n+\alpha\tanh(W\bm r_n+W_{\rm in}\bm u_n).
\label{eq:rc-training}
\end{equation}
Here, $\bm r_n$ is the reservoir state, $\bm u_n$ is the input, $\alpha$ is the leaking coefficient, $W_{\rm in}$ is the input matrix, and $W$ is the recurrent coupling matrix of the reservoir. To distinguish the intrinsic instability of the recurrent core from task-induced instability, we compare two architectures: an asymmetric random reservoir and a symmetric random reservoir. The latter provides a useful control because its isolated dynamics do not exhibit a conventional order--chaos transition over the parameter range studied here~\cite{Wang1998}. In both cases, $W$ is rescaled to have spectral radius $\rho$, which serves as the control parameter. The readout is $\bm y_n=W_{\rm out}\bm r_n$, with $W_{\rm out}$ trained from data. During autonomous forecasting, the output is fed back as input, yielding the closed-loop generative dynamics
\begin{equation}
\bm r_{n+1}=(1-\alpha)\bm r_n+\alpha\tanh[(W+W_{\rm in}W_{\rm out})\bm r_n].
\label{eq:closed}
\end{equation}
Setting $W_{\rm in}=0$ yields the isolated (bare) reservoir. Thus, for each target and architecture, the same recurrent core defines three operational regimes: isolated, teacher-forced, and closed-loop generative. We compute their maximal Lyapunov exponents $\Lambda$ as functions of $\rho$ and compare the resulting regime-dependent Lyapunov edges with the forecasting optimum obtained from direct performance scans. Details of the reservoir construction, RC implementation, Lyapunov-exponent calculations, and target models are provided in Supplementary Material (\SMsec{I} and \SMsec{II})~\cite{Supplemental}.

\begin{figure}[tbp]
\centering
\includegraphics[width=\linewidth]{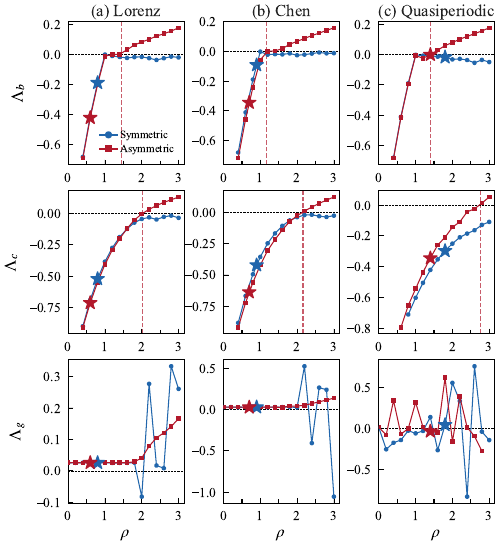}
\caption{Maximal Lyapunov exponents of the bare ($\Lambda_b$, top row), teacher-forced ($\Lambda_c$, middle row), and closed-loop generative ($\Lambda_g$, bottom row) reservoirs as functions of the spectral radius $\rho$ for the (a) Lorenz, (b) Chen, and (c) quasiperiodic targets. Red and blue curves correspond to asymmetric and symmetric reservoirs, respectively. Stars mark the forecasting-optimal spectral radii obtained from direct performance scans, and vertical dashed lines mark the Lyapunov edges of the asymmetric reservoirs.}
\vspace{-0.4cm}
\label{fig:regimes}
\end{figure}

Figure~\ref{fig:regimes} compares the three Lyapunov edges with the optimal radii obtained from direct performance scans. The top row shows the results for the bare reservoirs. For asymmetric reservoirs, the maximal Lyapunov exponent $\Lambda_b$ crosses zero as $\rho$ increases, reminiscent of the conventional order--chaos transition in random recurrent networks. This crossing is absent for symmetric reservoirs over the scanned range. In either case, the bare Lyapunov edge does not locate the forecasting optimum. The middle row shows the results for the teacher-forced reservoirs, whose stability is characterized by the conditional maximal Lyapunov exponent $\Lambda_c$~\cite{Ibanez2018,Lymburn2019,Lu2020,Platt2021,Hart2024}. For asymmetric reservoirs, $\Lambda_c$ also crosses zero, but at a substantially larger value of $\rho$ than the zero crossing of $\Lambda_b$. This shift is expected, as input driving improves the stability of the reservoir network~\cite{Rajan2010,Schuecker2018}. For symmetric reservoirs, no conditional Lyapunov edge appears within the scanned interval. Thus, the Lyapunov edge of the teacher-forced reservoir also fails to locate the forecasting optimum.

The bottom row shows the results for the closed-loop generative reservoirs. Here, the sign of the maximal Lyapunov exponent $\Lambda_g$ is not a universal criterion for optimality. For the chaotic Lorenz and Chen targets, faithful generation requires $\Lambda_g>0$, whereas $\Lambda_g<0$ indicates that the reservoir is collapsed to a steady state. For the quasiperiodic target, the desired long-time dynamics is nonchaotic, so a properly trained reservoir should have $\Lambda_g$ close to zero; positive $\Lambda_g$ indicates spurious chaos, whereas negative $\Lambda_g$ implies convergence to a fixed point. Taken together, Fig.~\ref{fig:regimes} shows that autonomous reservoir forecasting has no single privileged Lyapunov edge. The isolated, teacher-forced, and generative regimes define distinct dynamical objects, each with its own stability boundary and operational meaning. The symmetric reservoirs make this point especially clear: although a conventional Lyapunov edge is absent in both the isolated and teacher-forced regimes, these reservoirs can still be trained successfully. The failure of the edge-of-chaos heuristic is therefore conceptual rather than merely quantitative.

\emph{Direction-resolved finite-time Lyapunov spectra.---}
The failure of the regime-dependent Lyapunov edges to identify the optimal radius does not mean that Lyapunov stability is irrelevant to RC performance. Rather, it indicates that the maximal Lyapunov exponent alone is insufficient to characterize the forecasting potential of a given reservoir. In fact, this scalar quantity overlooks two essential features underlying successful RC. First, as a high-dimensional recurrent network, the reservoir responds to the input signal through the collective dynamics of its coupled units, whereas the maximal Lyapunov exponent captures only the dynamics along the most unstable tangent direction. Therefore, the target dynamics are represented not by the most unstable mode alone, but by a set of Lyapunov modes distributed across the spectrum. Second, the maximal Lyapunov exponent is a long-time average that suppresses temporal variations in local stability along the driven trajectory. During teacher-forced evolution, the reservoir undergoes both locally expanding and locally contracting episodes: the former amplify input-induced differences and enrich the reservoir representation, whereas the latter promote convergence, stability, and reproducibility. Both effects are essential for successful forecasting, but neither is resolved by the maximal Lyapunov exponent.

Motivated by these considerations, we characterize the teacher-forced reservoir using direction-resolved finite-time Lyapunov spectra~\cite{Rivkind2017,Engelken2023}. For a fixed spectral radius $\rho$, we first calculate the long-time-averaged conditional Lyapunov exponents and order them as $\lambda_1(\rho)\geq\lambda_2(\rho)\geq\cdots\geq\lambda_N(\rho)$, where $N$ is the reservoir dimension and $\lambda_1(\rho)=\Lambda_c(\rho)$. For the $i$th mode, we then calculate the finite-time exponent $\gamma_i(k;\rho)$ over the $k$th time window, such that $\lambda_i(\rho)=\langle\gamma_i(k;\rho)\rangle_k$. The averaged exponent $\lambda_i$ determines whether the mode is stable in the long-time sense, whereas the temporal fluctuations of $\gamma_i$ quantify how strongly its local stability is modulated by the target drive. To quantify these two properties, we define the stability margin of the $i$th mode as
\[m_i(\rho)=\max[0,-\lambda_i(\rho)],\]
and its fluctuation amplitude as
\[\sigma_i(\rho)=\mathrm{std}_k[\gamma_i(k;\rho)].\]
Thus, $m_i$ measures the long-time stable support of mode $i$, whereas $\sigma_i$ serves as a measure of its target-induced expressivity: the larger $m_i$ is, the more stable the mode is; the larger $\sigma_i$ is, the more strongly its local stability responds to the target drive. Details of the finite-time Lyapunov calculations are given in \SMsec{III}~\cite{Supplemental}.

\begin{figure}[tbp]
\centering
\includegraphics[width=\linewidth]{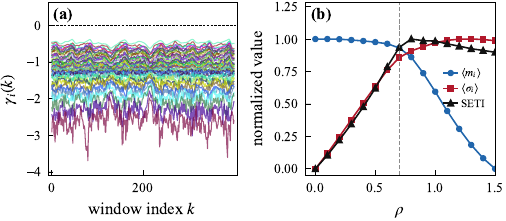}
\caption{(a) Temporal evolution of the finite-time Lyapunov exponents $\gamma_i(k)$ for selected representative modes in a teacher-forced Lorenz reservoir at a spectral radius of $\rho=0.7$. The modes are ordered by their long-time-averaged exponents $\lambda_i$ and selected from different regions of the Lyapunov spectrum. Although the driven reservoir is conditionally stable on average, these modes exhibit markedly different temporal fluctuations, revealing nonuniform target-induced modulation of local stability. (b) Normalized mean stability margin $\langle m_i\rangle$, mean fluctuation amplitude $\langle\sigma_i\rangle$, and stability--expressivity transfer index (\SETI) as functions of $\rho$. As $\rho$ increases, target-induced modulation becomes stronger while the stability margin decreases; \SETI{} peaks near the forecasting-optimal regime marked by the vertical dashed line.}
\vspace{-0.3cm}
\label{fig:ftle}
\end{figure}

Figure~\ref{fig:ftle}(a) shows the finite-time exponents of representative Lyapunov modes in a teacher-forced reservoir close to the forecasting optimum for the Lorenz oscillator. Although their long-time-averaged exponents are negative, indicating conditional stability of the driven reservoir, the modes exhibit strongly nonuniform temporal behavior. The displayed traces reveal a systematic trend: modes close to marginal stability fluctuate only weakly, whereas stable modes deeper in the spectrum exhibit pronounced temporal modulation; see \SMsec{III} for details~\cite{Supplemental}. This result suggests that target-induced responsiveness is not carried by fragile modes near the zero-exponent boundary, but is instead supported by stable modes whose local stability is strongly modulated by the target drive. Importantly, these stable modes combine two properties important for machine learning and forecasting: expressivity, because they respond strongly to the input, and stability, because their long-time-averaged exponents remain negative.

The finite-time Lyapunov spectra vary systematically with the spectral radius $\rho$. For small $\rho$, the reservoir is strongly contracting: the stability margins are large, but the finite-time fluctuations are weak. In this regime, the reservoir response is stable but insufficiently expressive. As $\rho$ increases, the target drive induces stronger modulation of the Lyapunov modes, thereby enhancing their expressivity, while the averaged spectrum approaches marginal stability and the stability margins decrease; see \SMsec{III} for further details~\cite{Supplemental}. The optimization of $\rho$ is therefore governed by a competition between target-induced expressivity and Lyapunov-mode stability. Figure~\ref{fig:ftle}(b) illustrates this competition more clearly: the normalized mean fluctuation amplitude $\langle\sigma_i\rangle$ increases with $\rho$, whereas the normalized mean stability margin $\langle m_i\rangle$ decreases. These opposing trends show that neither quantity alone can locate the forecasting optimum, calling for a dynamical criterion that balances the expressivity and stability of the Lyapunov modes.

\emph{Stability--expressivity transfer index.---} 
Motivated by the findings of direction-resolved finite-time Lyapunov analysis, we introduce the stability--expressivity transfer index (\SETI{}) to characterize the forecasting potential of a reservoir:
\begin{equation}
\SETI(\rho)=\frac{1}{N}\sum_{i=1}^{N}m_i(\rho)\sigma_i(\rho).
\label{eq:seti}
\end{equation}
Here, $m_i$ measures the long-time stability margin of the $i$th mode, whereas $\sigma_i$ characterizes its target-induced expressivity. The product $m_i\sigma_i$ assigns greater weight to modes that are both stable and expressive, while suppressing modes that possess only one of these properties. While this product form can be refined to better characterize the forecasting potential, it provides one of the simplest mode-level measures of the competition between stability and expressivity. Physically, \SETI{} quantifies the target-induced responsiveness supported by stable Lyapunov modes and assesses whether the dynamical structure created under driving can be retained during the transition from driven learning to autonomous generation. Note that \SETI{} is computed entirely from the teacher-forced reservoir and requires neither forecasting errors nor information from the generated trajectory.

\begin{figure}[tbp]
\centering
\includegraphics[width=\linewidth]{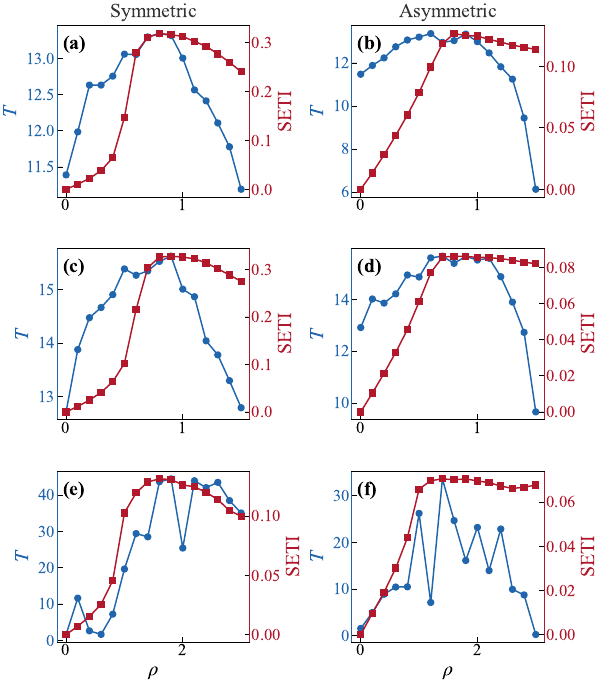}
\caption{Forecasting horizon $T$ obtained from autonomous forecasting (blue curves, left axes) and \SETI{} computed from teacher-forced reservoirs (red curves, right axes) as functions of $\rho$. The left and right columns correspond to symmetric and asymmetric reservoirs, respectively. The rows correspond to Lorenz, Chen, and quasiperiodic targets. Across all six target-architecture combinations, the maximum value of \SETI{} falls within the spectral-radius regime, where autonomous forecasting performs best.}
\vspace{-0.5cm}
\label{fig:prediction}
\end{figure}

The ability of \SETI{} to identify the optimal spectral radius is tested in Fig.~\ref{fig:prediction}. For each of the six target--architecture combinations, we compare \SETI{}$(\rho)$ with the forecasting horizon $T(\rho)$ obtained independently from closed-loop rollouts; the evaluation of $T(\rho)$ is described in \SMsec{IV}~\cite{Supplemental}. Although the two curves are not expected to overlap quantitatively over the scanned interval, they exhibit the same qualitative dependence on $\rho$: as $\rho$ increases, both \SETI{} and $T$ first increase and then decrease, reaching their maxima within an intermediate spectral-radius regime. This nonmonotonic trend is observed in all six cases shown in Fig.~\ref{fig:prediction}. Notably, the maximum of \SETI{} lies within, or very close to, the regime in which $T$ is largest. When the forecasting performance forms a broad plateau rather than a sharp peak, as in Figs.~\ref{fig:prediction}(b) and \ref{fig:prediction}(d), the relevant comparison is between optimal intervals rather than isolated points. The large fluctuations of $T$ in Figs.~\ref{fig:prediction}(e) and \ref{fig:prediction}(f) arise from the sensitivity of the closed-loop forecasting dynamics to variations in $\rho$, as discussed in \SMsec{IV}~\cite{Supplemental}. These results show that \SETI{}, although evaluated entirely from the teacher-forced dynamics, identifies the same operating regime as direct performance scans of autonomous forecasting.

Two important features of \SETI{} in Fig.~\ref{fig:prediction} are worth emphasizing. First, the agreement between \SETI{} and the forecasting horizon holds for both asymmetric and symmetric reservoirs, ruling out the possibility that the \SETI{} peak is merely a remnant of a conventional Lyapunov edge. For asymmetric reservoirs, the maximal Lyapunov exponents increase with $\rho$ [see Fig.~\ref{fig:regimes}], whereas \SETI{} varies nonmonotonically and peaks near the forecasting-optimal radius [see Fig.~\ref{fig:prediction}]. The distinction is even clearer for symmetric reservoirs: neither the isolated nor the teacher-forced Lyapunov spectrum exhibits a zero crossing within the scanned range, yet both the forecasting optimum and the \SETI{} maximum remain well defined. These results further confirm that the optimal radius is characterized not by the edge of chaos in the recurrent core, but by the regime in which target-induced expressivity is most effectively supported by stable Lyapunov modes. Second, the results for the quasiperiodic target show that \SETI{} is not tied to a particular type of generated dynamics [see Figs.~\ref{fig:prediction}(e) and \ref{fig:prediction}(f)]. In this case, the forecasting horizon fluctuates irregularly with $\rho$ because the closed-loop reservoir may drift toward an incorrect attractor or collapse to a steady state, as detailed in \SMsec{IV}~\cite{Supplemental}. By contrast, \SETI{} varies smoothly with $\rho$ and still identifies the best forecasting regime. Thus, across chaotic, quasiperiodic, and collapsed closed-loop dynamics, \SETI{} provides a unified indicator of forecasting potential. This robustness indicates that \SETI{} does not simply diagnose chaos, stability, or any particular long-term state of the forecasting reservoir. Rather, it measures the extent to which the target-induced dynamical structure acquired during teacher forcing can be transferred to autonomous generation.

\emph{Discussion.---} 
Our study sheds new light on the working mechanism of RC and suggests a mode-resolved framework for its design and optimization. Our study shows that a target is represented not by a single dominant direction, but by a broad set of Lyapunov modes with distinct dynamical roles. The main contribution comes from modes that remain stable on average while responding strongly to the input. Reservoir performance, therefore, reflects a balance between stability and expressivity. This stability--expressivity balance reformulates RC optimization as the search for reservoir configurations whose driven finite-time Lyapunov spectra support target-induced responsiveness in stable directions. The finite-time Lyapunov analysis provides not only a tool for understanding how RC works, but also a basis for developing new design strategies. In particular, reservoirs with the same global hyperparameters may perform very differently because their recurrent matrices organize the Lyapunov modes differently. This suggests tailoring the recurrent coupling matrix to the target dynamics and training the readout to favor stable and expressive directions. It may also relax the usual requirement that the bare reservoir should be operated close to a marginally stable regime, as even a high-dimensional or hyperchaotic network may serve as an effective reservoir if the target-relevant representation is supported by suitable stable modes. 

Several questions remain open. The role of the marginal modes near the top of the Lyapunov spectrum is not yet understood. One possibility is that these modes help generate or sustain the strong finite-time fluctuations observed in the deeper stable modes, but this conjecture requires a careful examination. \SETI{} should likewise be regarded as a first and minimal measure of stability--expressivity transfer. In its present form, it captures only the basic competition between mode stability and target-induced modulation, and its agreement with the forecasting horizon is therefore mainly qualitative. More refined measures may incorporate nonlinear weighting, correlations among modes, or higher-order mode interactions. Finally, whether the same stability--expressivity transfer picture can guide the optimization of other hyperparameters remains an important question.

The source code and data supporting the results of this study are publicly available in Ref.~\cite{Github}.

\begin{acknowledgments}
This work was supported by the National Natural Science Foundation of China (Grant No.~12275165).
\end{acknowledgments}

\end{document}